\documentclass[conference]{IEEEtran}
\IEEEoverridecommandlockouts

\usepackage{cite}
\usepackage{amsmath,amssymb,amsfonts}
\usepackage{graphicx}
\usepackage{textcomp}
\usepackage{xcolor}
\def\BibTeX{{\rm B\kern-.05em{\sc i\kern-.025em b}\kern-.08em
    T\kern-.1667em\lower.7ex\hbox{E}\kern-.125emX}}

\usepackage{physics}
\usepackage{hyperref}
\usepackage{tikz}
\usetikzlibrary{arrows.meta,positioning}
\usepackage{algorithm}
\usepackage{algpseudocode} 
\usepackage{booktabs}

\begin{document}

\title{Bayesian Sequential Verification for Budget-Aware Quantum Program Testing}

\author{\IEEEauthorblockN{Lei Zhang}
\IEEEauthorblockA{\textit{Department of Information Systems} \\
\textit{University of Maryland, Baltimore County}\\
Maryland, USA \\
leizhang@umbc.edu}
}

\maketitle

\begin{abstract}

Quantum programs often produce probability distributions rather than deterministic outputs, making verification inherently statistical and increasingly costly on real hardware. In practice, developers still frequently rely on testing with fixed shot budgets on simulators, which are simple but time-consuming and poorly suited to noisy backends. What is missing is a verification approach that is both statistically explicit and budget-aware.

This paper formulates Bayesian sequential verification as a reference-based Bayesian hypothesis testing workflow in which priors are derived from explicit reference sources, such as finite-shot reference runs or ideal/statevector-based computation, and verification decisions are updated batch by batch as measurement evidence accumulates. 

This approach is evaluated in Qiskit on two complementary workloads: Bell-state and QAOA-MaxCut. Across both case studies, the results show that Bayesian sequential verification can substantially reduce measurement costs compared to fixed-budget baselines when the success probability of the program exceeds the target threshold. 

The findings position Bayesian sequential verification as a practical verification workflow for quantum programs. The approach provides a foundation for future quantum continuous-integration pipelines that require reliable, budget-aware pass/fail decisions and motivates validation on real quantum hardware.

\end{abstract}


\section{Introduction}
\label{sec:intro}

Quantum programs increasingly produce probability distributions rather than single deterministic outputs, relying on repeated measurement to estimate performance-relevant properties~\cite{zhao2020quantum,miranskyy2019testing,abreu2026softwaretestingquantumworld}. Unlike classical probabilistic programs, they do not expose an internal state that can be deterministically checked at runtime~\cite{miranskyy2020your,kim2025detecting}; instead, correctness must be inferred from stochastic outcomes whose distribution depends on quantum hardware noise, compilation choices, and device drift~\cite{zhang2023identifying,virani2025distinguishing}. As a result, verification is inherently distributional and resource constrained: each additional shot consumes real hardware budget (e.g., IBM quantum resources can cost \$96 USD per minute)~\cite{awschalom2025challenges}, outcomes may shift over time, and practically meaningful criteria are often task-level (e.g., success probability over a set of high-quality solutions) rather than exact functional equivalence on individual inputs. 

Despite this, current practice often relies on fixed and ad-hoc shot budgets (e.g., 10k shots per circuit)~\cite{dutta2018testing}. Such fixed-shot testing is easy to apply, but statistically inefficient. If the observed behavior of the program (e.g., its success probability) is clearly above the target threshold, a large fixed budget can waste shots. Otherwise, the same budget may be insufficient because sampling uncertainty can be too large to support a confident decision. 

To address these limitations, this paper formulate a Bayesian sequential verification approach for quantum programs: a reference-based Bayesian Hypothesis Testing (BHT)~\cite{bernardo2002bayesian} workflow in which the prior is derived from an explicit reference source (e.g., shot-derived from a noiseless simulator, or a statevector distribution), while verification decisions are updated batch by batch as shots are collected rather than committed to a fixed budget in advance. Concretely, this paper casts verification as deciding whether a backend’s property success probability exceeds a target threshold, and we implement a sequential Bayesian procedure that updates a Beta posterior from batched measurements and verifies once a conservative lower credible bound (LCB) crosses the target. To summarize, this approach encodes prior knowledge from simulation or prior runs, quantifies uncertainty under finite shots, and expresses verification outcomes as explicit evidence thresholds. These properties make BHT a promising foundation for future quantum continuous-integration (CI) settings, where verification must be both budget-aware and statistically defensible.

To validate the BHT workflow, this paper implements a Qiskit-based~\cite{mckay2018qiskit} harness and applies it on two case studies: (i) Bell-state~\cite{kim2001quantum} correlation verification in complementary $Z$/$X$ measurement bases~\cite{nielsen2010quantum}, and (ii) the Quantum Approximate Optimization Algorithm (QAOA)~\cite{choi2019tutorial} for combinatorial optimization (e.g., MaxCut~\cite{goemans1995improved}) using a Top-$K$ success predicate under ideal and noisy execution models~\cite{zhou2020quantum}. In both cases, we also compare BHT against a matched fixed-budget baseline to assess whether the observed shot savings persist beyond cap-relative comparisons.

The \textbf{contributions} of this paper are: (i) formulating a Bayesian sequential verification procedure for quantum programs under finite shot budgets; (ii) evaluating this workflow on two complementary Qiskit case studies across multiple IBM fake backends with noise, including matched fixed-budget baseline~\cite{ibm2026quantum}, and 3) releasing the artifacts publicly on Zenodo for reproduction and reuse at \href{https://doi.org/10.5281/zenodo.18341209}{\nolinkurl{doi:10.5281/zenodo.18341209}}.

Note that this paper does not claim a new Bayesian statistical method; rather, it operationalize BHT as a reusable verification workflow for quantum programs under finite shot budgets. Overall, this research aims to make quantum regression checks more budget-aware and statistically rigorous by replacing fixed-shot testing with explicit posterior guarantees. Moreover, the case studies are  initial evidence of feasibility for future CI-driven quantum regression testing.

The remainder of this paper is organized as follows. Section~\ref{sec:background} motivates Bayesian sequential verification in the context of quantum-program testing. Section~\ref{sec:method} presents the reference-based BHT workflow, including prior construction, posterior updates, and stopping rules. Section~\ref{subsec:case1-bell} evaluates the approach on Bell-state verification, and Section~\ref{subsec:case2-qaoa} studies a more challenging QAOA Top-$K$ verification task. Section~\ref{sec:discussion} discusses the baseline comparison and summarizes lessons from the matched fixed-budget baselines. Finally, Sections~\ref{sec:threats} and~\ref{sec:conclusion} discuss threats to validity and conclude with directions for future work.

\section{Background and Motivation}
\label{sec:background}

In classical verification, Bayesian statistical model checking and sequential decision procedures provide principled mechanisms to accept, reject, or continue testing as observations accrue~\cite{zuliani2013bayesian,erdogmus2022bayesian}.
However, many classical applications assume relatively stable observation processes and specifications evaluated over deterministic traces or well-controlled stochastic models.

Quantum software pushes these assumptions: observable behavior is sample-based, hardware is costly, and nonstationary noise can cause statistically meaningful variation even when the source code is unchanged. Recent work has begun to characterize noise and detect flakiness in quantum programs, underscoring the need for verification methods that are distribution-aware and can separate genuine regressions from hardware-induced variance~\cite{zhang2023identifying,virani2025distinguishing}. Moreover, Miranskyy’s work~\cite{miranskyy2025cost} develops a unified framework for estimating the number of measurements required for quantum program verification, analyzes inverse, swap, and chi-square tests, and studies how verification budgets scale at the program level. 
Our work differs from the above literature, unlike estimating the measurement cost of fixed testing procedures, we formulate a sequential Bayesian decision process that can stop early once sufficient posterior evidence has accumulated. 

In quantum advantage validation---including boson sampling~\cite{spring2013boson} and random-circuit-style experiments~\cite{bouland2019complexity}---Bayesian hypothesis testing and likelihood-based model comparison have been used to contrast a target quantum model against classical or adversarial alternatives, reporting evidential measures such as posterior odds or Bayes factors~\cite{martinez2023classical,martinez2024linear}. These efforts demonstrate two properties that are particularly relevant to near-term quantum computing: (i) evidence can be accumulated sequentially as samples arrive, and (ii) prior knowledge can be represented explicitly rather than remaining informal.

Our setting differs from prior quantum-validation use cases in both goal and workflow. This paper focuses on quantum program verification for software engineering: enabling repeatable checks such as regression testing across runs, backends, transpiler settings, and library updates, rather than demonstrating quantum advantage. In practice, quantum developers often have usable reference information, such as small-instance simulations used during unit testing, previously certified baselines from nightly builds, and historical test artifacts collected by CI pipelines. Yet this reference information is rarely integrated into test decisions in a statistically explicit and configurable way. Our contribution is to treat these artifacts as explicit reference sources for prior construction and to show how Bayesian sequential verification can use them to guide sample-efficient regression checks while preserving conservative decision guarantees.

\section{Bayesian Sequential Verification for Quantum Programs}
\label{sec:method}

\begin{figure}[t]
    \centering
    \includegraphics[width=0.9\linewidth]{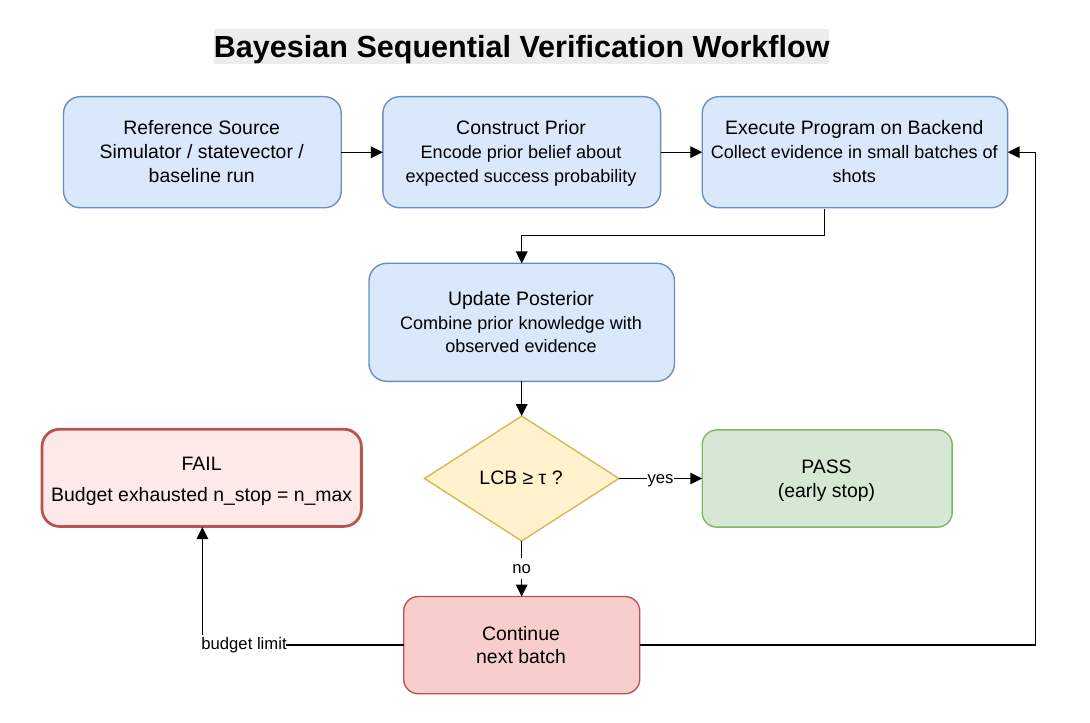}
    \caption{Overall illustration of the Bayesian sequential verification workflow.}
    \label{fig:bht-workflow}
\end{figure}

A quantum program (circuit) $P$ executed on a backend $B$ induces an unknown outcome distribution $p_B$ over a discrete space $\mathcal{Y}$ (e.g., bitstrings). Each execution yields an outcome $y\in\mathcal{Y}$. This paper assumes that, within a single run, outcomes are conditionally i.i.d.\ given $p_B$ and that $p_B$ is stationary over the shots collected in that run.\footnote{We use independent seeds and short runs to reduce the risk of long-term drift; handling drift explicitly is outside the scope of this paper.}

Both case studies instantiate verification as a property-based Bernoulli predicate $\phi:\mathcal{Y}\rightarrow\{0,1\}$ (success/failure), inducing an unknown backend success probability $\theta:=\Pr_{y\sim p_B}(\phi(y)=1)$. 
As shown in Figure~\ref{fig:bht-workflow}, this paper performs Bayesian sequential verification by (i) initializing a prior for $\theta$ from a reference success rate $\mu_0$ and prior strength $s$, (ii) updating the posterior as shots are collected in batches, and (iii) declaring \textsf{PASS} once a conservative one-sided LCB meets the target threshold under a stopping rule. 

\begin{algorithm}[t]
\caption{Sequential Bayesian verification for a property $\phi$}
\label{alg:seq-bht}
\begin{algorithmic}[1]
\Statex \textbf{Inputs:} program $P$, backend $B$, outcome space $\mathcal{Y}$
\Statex \hspace{\algorithmicindent} specification $\phi:\mathcal{Y}\rightarrow\{0,1\}$
\Statex \hspace{\algorithmicindent} prior source $\mathcal{R}$ (shot-derived or analytic/statevector-derived)
\Statex \hspace{\algorithmicindent} prior strength $s>0$, targets $(\tau,\delta)$, batch size $b$
\Statex \hspace{\algorithmicindent} shot cap $n_{\max}$, minimum-shot gate $n_{\min}$, pass-streak requirement $r$
\Statex \textbf{Outputs:} decision \textsf{PASS}/\textsf{FAIL}, stopping time $n_{\mathrm{stop}}$

\State Compute reference success rate $\mu_0 \gets \Pr_{\mathcal{R}}(\phi(y)=1)$
\State Set prior parameters: $\alpha_0 \gets 1+s\mu_0$, $\beta_0 \gets 1+s(1-\mu_0)$
\State Initialize $n\gets 0$, $k\gets 0$, $streak\gets 0$
\While{$n < n_{\max}$}
  \State Execute $P$ on $B$ for $b$ shots and obtain outcomes $y_{1:b}$
  \State $k \gets k + \sum_{j=1}^{b}\phi(y_j)$; \quad $n \gets n + b$
  \State Compute posterior lower bound $\mathrm{LCB} \gets \mathrm{BetaInv}\bigl(\delta;\,\alpha_0+k,\;\beta_0+n-k\bigr)$
  \If{$n \ge n_{\min}$ \textbf{and} $\mathrm{LCB} \ge \tau$}
    \State $streak \gets streak + 1$
    \If{$streak \ge r$}
      \State \Return \textsf{PASS}, $n_{\mathrm{stop}} \gets n$
    \EndIf
  \Else
    \State $streak \gets 0$
  \EndIf
\EndWhile
\State \Return \textsf{FAIL}, $n_{\mathrm{stop}} \gets n_{\max}$
\end{algorithmic}
\end{algorithm}

\subsection{Prior construction}
\label{subsec:priors}

In this paper, prior information supports two reference sources, i.e., shot-derived priors and computed priors. In both cases, we denote the reference success rate by $\mu_0$.

\textbf{Shot-derived priors (empirical reference).} 
When a measurement-driven reference workflow is preferred, we estimate the reference success rate from a noiseless or baseline backend $B_{\mathrm{ref}}$ by running $P$ for $n_0$ shots:
\[
\mu_0 \;=\; \frac{k_0}{n_0},\qquad
k_0=\sum_{i=1}^{n_0}\phi(y_i^{(\mathrm{ref})}).
\]
This is the approach used in the Bell-state case study (Section~\ref{subsec:case1-bell}).

\textbf{Computed priors (analytic/statevector reference).}
When the reference success probability can be computed directly, we set
\[
\mu_0 \;=\; \Pr_{\mathrm{ref}}\bigl(\phi(y)=1\bigr)
      \;=\; \sum_{y\in\mathcal{Y}} \phi(y)\,p_{\mathrm{ref}}(y),
\]
which avoids additional Monte Carlo variance in the prior.
This is the approach used in the QAOA case study (Section~\ref{subsec:case2-qaoa}).

In either case, we model the success probability $\theta$ using a Beta prior, which is the conjugate prior for Bernoulli observations and enables closed-form posterior updates. We instantiate the prior as
\begin{equation}
\label{eq:beta-prior}
\theta \sim \mathrm{Beta}(\alpha_0,\beta_0),\qquad
\alpha_0 = 1 + s\,\mu_0, \quad \beta_0 = 1 + s\,(1-\mu_0),
\end{equation}
where $\mu_0$ is the reference success rate and $s$ controls the prior strength. We interpret $s$ as an equivalent prior sample size, so that the prior contributes $s$ pseudo-observations centered at $\mu_0$, regularizing early batches without dominating the posterior once hundreds to thousands of backend shots accumulate. Empirically, we set the prior strength to $s=20$ in all experiments.

\subsection{Posterior updates}
\label{subsec:posterior}

Program $P$ is executed on the target backend $B$ in batches of size $b$.
After batch $t$, let $n_t$ be the cumulative number of shots and $k_t$ the cumulative number of successes under $\phi$.
The posterior is
\begin{equation}
\label{eq:beta-posterior}
\theta \mid (k_t,n_t) \sim \mathrm{Beta}(\alpha_0+k_t,\ \beta_0+n_t-k_t).
\end{equation}
This update can be performed online after each batch and directly supports the one-sided LCB used by the stopping rule in Section~\ref{subsec:decision}.

\subsection{Decision rule and stopping policy}
\label{subsec:decision}

Given a target threshold $\tau\in(0,1]$ and tail probability $\delta$, we compute a conservative one-sided LCB
\begin{equation}
\label{eq:lcb}
\mathrm{LCB}_{1-\delta}(\theta)
= \mathrm{BetaInv}\bigl(\delta;\ \alpha_0+k,\ \beta_0+n-k\bigr).
\end{equation}
After each batch, once the cumulative shots satisfy $n\ge n_{\min}$, we check whether
\[
\mathrm{LCB}_{1-\delta}(\theta)\ge \tau.
\]
This algorithm stops and outputs \textsf{PASS} at the earliest stopping time $n_{\mathrm{stop}}$ for which this condition holds for $r$ consecutive batch checks (the pass-streak requirement). If the shot cap $n_{\max}$ is reached before this occurs, the algorithm stops and outputs \textsf{FAIL} with $n_{\mathrm{stop}}=n_{\max}$.

\subsection{Evaluation metrics}
\label{subsec:evaluation}

The verifier is evaluated over repeated randomized runs (different seeds) to characterize both decision behavior and sample efficiency.
For each configuration (backend $B$ and threshold $\tau$), we execute $R$ independent runs and report:
\begin{enumerate}
    \item PASS rate, the fraction of runs that satisfy the stopping rule before reaching $n_{\max}$;
    \item shot cost, the stopping time $n_{\mathrm{stop}}$ (set to $n_{\max}$ for \textsf{FAIL} runs); and
    \item dispersion, summarized by the interquartile range (IQR) of $\{n_{\mathrm{stop}}\}$ reported as $[Q_1,Q_3]$.
\end{enumerate}

\subsection{Experimental settings}
\label{subsec:settings}

Unless otherwise noted, both case studies instantiate Algorithm~\ref{alg:seq-bht} with the same sequential verification policy: batch size $b=50$, pass-streak requirement $r=2$, and prior strength $s=20$. For Bell-state verification, the shot cap is $n_{\max}=2{,}000$ and the minimum-shot gate is $n_{\min}=200$. For QAOA verification, the shot cap is $n_{\max}=10{,}000$ and the minimum-shot gate is $n_{\min}=1{,}000$. Thus, the two case studies differ only in the success predicate $\phi$ and in how the reference success rate $\mu_0$ is constructed: Bell uses a shot-derived reference from a noiseless simulator, whereas QAOA uses a statevector-derived ideal reference.

\section{Use Case 1: Bell state ($Z$ \& $X$ bases)}
\label{subsec:case1-bell}

\begin{table*}[t]
\centering
\small
\caption{Bell-state verification across IBM fake backends using a noiseless prior (\texttt{AerSimulator}). Each \(\tau\) uses \(R{=}10\) seeds; \textsf{PASS} requires both \(Z\)- and \(X\)-basis checks to verify within a shot cap of \(2{,}000\). This table reports the median per-run cost \(\mathrm{median}(\max(n_Z,n_X))\) and IQR \([Q_1,Q_3]\).}
\label{tab:bell_bht_three_backends}
\begin{tabular}{c|ccc|ccc|ccc}
\hline
 & \multicolumn{3}{c|}{\texttt{FakeNairobiV2}} & \multicolumn{3}{c|}{\texttt{FakeLimaV2}} & \multicolumn{3}{c}{\texttt{FakeManilaV2}} \\
\(\tau\) & PASS & Med. max-shots & IQR & PASS & Med. max-shots & IQR & PASS & Med. max-shots & IQR \\
\hline
0.85 & 10/10 & 250  & [250, 250]  & 10/10 & 250  & [250, 250]  & 10/10 & 250  & [250, 250] \\
0.88 & 10/10 & 400  & [288, 488]  & 10/10 & 250  & [250, 250]  & 10/10 & 250  & [250, 250] \\
0.90 &  8/10 & 1175 & [813, 1738] & 10/10 & 325  & [250, 400]  & 10/10 & 350  & [250, 400] \\
0.92 &  0/10 & 2000 & [2000, 2000] & 10/10 & 575  & [250, 850]  & 10/10 & 825  & [350, 1250] \\
0.94 &  0/10 & 2000 & [2000, 2000] &  3/10 & 2000 & [2000, 2000] &  0/10 & 2000 & [2000, 2000] \\
\hline
\end{tabular}
\end{table*}

\textbf{Goal.}
This paper studies a 2-qubit circuit that prepares the Bell state
\[
\ket{\Phi^{+}}=(\ket{00}+\ket{11})/\sqrt{2}
\]
by applying a Hadamard gate to qubit 0 followed by a CNOT from qubit 0 to qubit 1.
In the ideal setting, measurement in the computational ($Z$) basis yields only outcomes $\{00,11\}$ with equal probability:
\[
P_{\text{ideal}}^{(Z)}=\{00:0.5,\,11:0.5\}.
\]
In practice, noise can introduce (i) population errors that leak probability mass to $\{01,10\}$ and/or (ii) phase decoherence that preserves $Z$-basis correlations while degrading coherence. To detect both effects, the state is verified in two complementary measurement configurations: the $Z$ basis (direct measurement after Bell-state preparation) and the $X$ basis, implemented by applying $H^{\otimes 2}$ before measurement. For $\ket{\Phi^{+}}$, the ideal $X$-basis outcomes are again perfectly correlated, so only $\{00,11\}$ appear in the post-Hadamard measurement frame.

\textbf{Bayesian sequential verification (per basis).}
For each basis $b\in\{Z,X\}$, a Bernoulli success event is defined as observing a correlated outcome:
\[
\begin{aligned}
\text{Success in } Z &: \; y \in \{00,11\}, \\
\text{Success in } X &: \; y \in \{00,11\}
\quad \text{after applying } H^{\otimes 2}.
\end{aligned}
\]
The prior is derived from shot-based estimates on a noiseless \texttt{AerSimulator} (2{,}000 shots per basis), yielding reference success rates $\mu_Z$ and $\mu_X$. The posterior is then updated sequentially using measurements collected from noisy IBM fake backends, with a per-run shot cap of $n_{\max}=2{,}000$. Thus, the testing procedure uses a fixed budget as an upper bound, while allowing early stopping when sufficient evidence is obtained.

\textbf{Results and interpretation.}
Table~\ref{tab:bell_bht_three_backends} reports Bell-state verification outcomes across three IBM fake backends for a sweep of thresholds \(\tau\) (10 seeds per setting). The results show a clear tradeoff: moderate thresholds \(\tau\) typically verify with significantly fewer than \(2{,}000\) shots, whereas stricter thresholds \(\tau\) increase both the median shot cost (sometimes running to the maximum number of shots) and its variability.

We highlight two observations. First, when the observed success probability \(\theta\) is well above the threshold \(\tau\), the posterior quickly accumulates strong evidence, so the test can stop early. For example, at \(\tau=0.85\), all three backends achieve \(\textsf{PASS}=10/10\) with a median of only \(250\) max-shots and a tight IQR \([250,250]\), corresponding to an \(87.5\%\) shot reduction relative to the fixed budget. Second, when the backend is near the decision boundary---often dominated by the more demanding coherence check in the \(X\) basis---the method becomes conservative, requiring more shots or returning a budget-limited \textsf{FAIL}. For example, at \(\tau=0.92\), \texttt{FakeNairobiV2} fails in all runs (\(0/10\)) and reaches the shot cap, whereas \texttt{FakeLimaV2} and \texttt{FakeManilaV2} still pass (\(10/10\)) with shot reductions of \(71.25\%\) and \(58.75\%\), respectively; their wider IQRs reflect increased variability near the threshold. These results suggest that practical regression workflows may need backend-specific calibration of \(\tau\) to balance assurance against test cost.

\section{Use Case 2: QAOA (Top-$K$)}
\label{subsec:case2-qaoa}

\begin{table*}[t]
\centering
\small
\caption{QAOA Top-$K$ verification across IBM backends using an ideal (statevector-derived) prior $\mu_0$ and a posterior-only LCB stopping rule. Each \(\tau\) uses \(R{=}10\) seeds. This table reports median stopping shots and IQR \([Q_1,Q_3]\); the shot cap is \(n_{\max}=10{,}000\).}
\label{tab:qaoa_topk_bht}
\begin{tabular}{c|ccc|ccc|ccc}
\hline
 & \multicolumn{3}{c|}{\texttt{AerSimulator}} & \multicolumn{3}{c|}{\texttt{FakeBrooklynV2}} & \multicolumn{3}{c}{\texttt{FakeKolkataV2}} \\
\(\tau\) & PASS & Med. shots & IQR & PASS & Med. shots & IQR & PASS & Med. shots & IQR \\
\hline
0.60 & 10/10 & 150 & [100, 188] &  9/10 & 650 & [250, 3850] & 10/10 & 1150 & [612, 4538] \\
0.70 & 10/10 & 225  & [112, 375]   &  7/10 & 4600  & [1138, 9962]  &  6/10 & 4050  & [1475, 10000] \\
0.80 & 10/10 & 375  & [212, 538]   &  5/10 & 9275  & [2750, 10000]  &  2/10 & 10000 & [10000, 10000] \\
0.85 & 10/10 & 450  & [262, 1138]  &  2/10 & 10000 & [10000, 10000] &  2/10 & 10000 & [10000, 10000] \\
0.90 & 10/10 & 525  & [375, 1875]  &  2/10 & 10000 & [10000, 10000] &  1/10 & 10000 & [10000, 10000] \\
0.95 &  8/10 & 2350 & [662, 7100]  &  2/10 & 10000 & [10000, 10000] &  1/10 & 10000 & [10000, 10000] \\
\hline
\end{tabular}
\end{table*}

\textbf{Goal.}
This paper studies an 8-qubit QAOA-based MaxCut instance and verifies whether the circuit produces high-quality solutions with sufficient frequency. Rather than verifying a single bitstring, we verify Top-$K$ success: the probability mass assigned to a target set $S_K$ containing the $K$ highest-scoring bitstrings under the objective. This metric is aligned with near-term optimization practice, where recovering any of several high-quality solutions is acceptable~\cite{wang2024red}.

\textbf{Top-$K$ success event.}
Let $z\in\{0,1\}^n$ denote a measured bitstring and let $C(z)$ be its MaxCut value. We define the success set $S_K$ as the $K$ highest-scoring bitstrings under the \emph{ideal} noiseless model and use the Bernoulli success predicate
\[
\phi(z)=\mathbb{I}[z\in S_K].
\]
The corresponding reference success probability is
\[
\mu_0=\Pr_{\mathrm{ideal}}(z\in S_K),
\]
which is computed directly from the ideal distribution and use to instantiate the Beta prior in Eq.~\ref{eq:beta-prior}.

A fixed MaxCut ring instance is used with \(n=8\) vertices and edge set
\(\{(0,1),(1,2),(2,3),(3,4),(4,5),(5,6),(6,7),(7,0)\}\), with QAOA depth \(p=1\).
We set \(K=5\) and select the angles \((\gamma^\star,\beta^\star)\) by a coarse ideal-statevector grid search over
\(\gamma\in\texttt{linspace}(0.1,2.9,15)\) and \(\beta\in\texttt{linspace}(0.1,1.5,15)\), yielding
\((\gamma^\star,\beta^\star)=(2.300,1.200)\).
For this instance and parameter setting, the ideal Top-5 success mass is \(\mu_0 \approx 0.1813\).

Then, Algorithm~\ref{alg:seq-bht} is applied on three target backends: \texttt{AerSimulator} (noiseless baseline), \texttt{FakeBrooklynV2}, and \texttt{FakeKolkataV2}.
For each $(\tau,\text{backend})$, we run $R{=}10$ seeds and report \textsf{PASS} rate and the distribution of stopping shots.

\textbf{Results and interpretation.}
Table~\ref{tab:qaoa_topk_bht} reports sequential Top-5 verification outcomes.

On \texttt{AerSimulator}, all runs verify for $\tau\le 0.90$ (\textsf{PASS} rate $=1.00$) with only a few hundred shots (median $150$--$525$). At $\tau=0.95$, verification becomes less stable: \textsf{PASS} rate drops to $0.80$, and shot cost increases sharply (median $2{,}350$, IQR $[662,7{,}100]$), indicating a near-threshold case.

On noisy \texttt{FakeBrooklynV2} and \texttt{FakeKolkataV2}, verification is already costly at $\tau=0.70$ (\textsf{PASS} $7/10$ and $6/10$, respectively), with medians in the $4{,}000$--$5{,}000$-shot range and wide variability. For $\tau\ge 0.80$, most runs become budget-limited at $10{,}000$ shots, suggesting that the proposed algorithm does not have enough evidence to pass, even after using the full budget.

Compared with the Bell-state case, QAOA verification is significantly more noise-sensitive: under hardware-realistic noise, both \textsf{PASS} probability and stopping-time stability degrade rapidly as $\tau$ increases. This reflects the fact that Top-$K$ success is a lower-probability event than Bell-state correlation, making the posterior lower bound slower to separate from the target under noise. These findings suggest that future Top-$K$ regression policies may benefit from more robust predicate design (e.g., larger $K$), multi-threshold stopping rules, and backend-specific calibration.

\section{Discussion}
\label{sec:discussion}

\begin{table*}[t]
\centering
\small
\caption{BHT vs. matched fixed-budget baselines across both case studies. For Bell-state verification, the baseline is a one-sided fixed-budget binomial LCB test using the same correlated-outcome success event and the same confidence target as BHT. For QAOA verification, the baseline uses the same success event, the same confidence target, and the same scaled decision target as BHT, i.e., \(\mathrm{LCB}_{0.975}(\theta)\ge \tau\mu_0\), but evaluates the rule only at the fixed budget. Each $\tau$ uses $R{=}10$ seeds. This paper reports BHT PASS count, fixed-baseline PASS count, and median shot saving of BHT relative to the fixed-budget baseline.}
\label{tab:baseline_comparison_all}
\resizebox{\textwidth}{!}{%
\begin{tabular}{c|ccc|ccc|ccc}
\toprule
\multicolumn{10}{c}{\textbf{Bell-state verification}} \\
\midrule
 & \multicolumn{3}{c|}{\texttt{FakeNairobiV2}} & \multicolumn{3}{c|}{\texttt{FakeLimaV2}} & \multicolumn{3}{c}{\texttt{FakeManilaV2}} \\
\(\tau\) & BHT PASS & Fixed PASS & Med. saving & BHT PASS & Fixed PASS & Med. saving & BHT PASS & Fixed PASS & Med. saving \\
\midrule
0.85 & 10/10 & 10/10 & 1750 & 10/10 & 10/10 & 1750 & 10/10 & 10/10 & 1750 \\
0.88 & 10/10 & 10/10 & 1600 & 10/10 & 10/10 & 1750 & 10/10 & 10/10 & 1750 \\
0.90 &  8/10 &  6/10 &  825 & 10/10 & 10/10 & 1675 & 10/10 & 10/10 & 1650 \\
0.92 &  0/10 &  0/10 &    0 & 10/10 & 10/10 & 1425 & 10/10 & 10/10 & 1175 \\
0.94 &  0/10 &  0/10 &    0 &  3/10 &  2/10 &    0 &  0/10 &  0/10 &    0 \\
\midrule
\multicolumn{10}{c}{\textbf{QAOA Top-$K$ verification}} \\
\midrule
 & \multicolumn{3}{c|}{\texttt{AerSimulator}} & \multicolumn{3}{c|}{\texttt{FakeBrooklynV2}} & \multicolumn{3}{c}{\texttt{FakeKolkataV2}} \\
\(\tau\) & BHT PASS & Fixed PASS & Med. saving & BHT PASS & Fixed PASS & Med. saving & BHT PASS & Fixed PASS & Med. saving \\
\midrule
0.60 & 10/10 & 10/10 & 9850 &  9/10 &  8/10 & 9350 & 10/10 &  9/10 & 8850 \\
0.70 & 10/10 & 10/10 & 9775 &  7/10 &  5/10 & 5400 &  6/10 &  3/10 & 5950 \\
0.80 & 10/10 & 10/10 & 9625 &  5/10 &  2/10 &  725 &  2/10 &  0/10 &    0 \\
0.85 & 10/10 & 10/10 & 9550 &  2/10 &  0/10 &    0 &  2/10 &  0/10 &    0 \\
0.90 & 10/10 & 10/10 & 9475 &  2/10 &  0/10 &    0 &  1/10 &  0/10 &    0 \\
0.95 &  8/10 &  7/10 & 7650 &  2/10 &  0/10 &    0 &  1/10 &  0/10 &    0 \\
\bottomrule
\end{tabular}%
}
\end{table*}

Table~\ref{tab:baseline_comparison_all} compares BHT against a matched fixed-budget baseline in both workloads. In Bell-state verification, the baseline evaluates the same basis-specific Bernoulli event---observing a correlated outcome---and the same confidence target, but makes its decision only at the fixed budget. In QAOA verification, the comparison is also matched at the level of the success event, confidence level, and decision target: both methods test whether \(\mathrm{LCB}_{0.975}(\theta)\ge \tau\mu_0\), where \(\theta\) is the Top-$K$ success probability, \(\mu_0\approx 0.1813\) is the ideal Top-$K$ success mass, and the fixed baseline evaluates this rule only at \(n_{\max}=10{,}000\). For QAOA, a fixed 8-node ring MaxCut instance is used with \(n=8\), QAOA depth \(p=1\), \(K=5\), batch size \(b=50\), pass-streak requirement \(r=2\), prior strength \(s=20\), minimum-shot gate \(n_{\min}=1{,}000\), and \(R=10\) seeds per threshold. The QAOA angles are selected once by coarse ideal-statevector grid search, yielding \((\gamma^\star,\beta^\star)=(2.300,1.200)\).

The baseline comparison shows that the advantage of the proposed method is not that BHT stops earlier than a configured shot cap, but that it can preserve a clear confidence rule. The sequential rule often achieves the same or higher PASS count as the fixed-budget baseline while using fewer shots. For example, this pattern is visible in Bell-state verification at moderate thresholds such as \(\tau=0.85\) and \(\tau=0.88\), where all three backends achieve identical PASS counts under both methods but BHT saves roughly \(1{,}600\)--\(1{,}750\) shots. A similar behavior appears in QAOA at lower thresholds: on \texttt{FakeBrooklynV2} at \(\tau=0.70\), e.g., BHT improves PASS from \(5/10\) to \(7/10\) while saving a median of \(5{,}400\) shots relative to the fixed-budget test.

As the target threshold approaches the effective success probability under noise, posterior lower bounds tighten more slowly and both sequential and fixed-budget rules become increasingly conservative. In Bell-state verification, this appears at thresholds such as \(\tau=0.92\) and \(\tau=0.94\), where savings can disappear on the more challenging backends. In QAOA, the effect is stronger because Top-$K$ success is a substantially smaller-probability event than Bell-state correlation. The matched baseline confirms this behavior: BHT provides clear gains at \(\tau=0.60\) and \(\tau=0.70\), more modest gains at \(\tau=0.80\), and no median-shot advantage once runs become budget-limited. 

As shown by the experiments and findings, the two case studies are complementary. Bell-state verification serves as a simple demonstration of the method: the success predicate is simple, and the ideal success probability is high. QAOA verification plays a different role as a stress test for the method, where early stopping remains effective and where noise pushes the problem into a near-threshold zone. 

These observations suggest a practical role for Bayesian sequential verification in future quantum CI workflows. When a system is well above specification, sequential verification can provide large-shot savings with little loss. When the system is near the decision boundary, however, both BHT and fixed-budget baselines become conservative, indicating the need for more robust mechanisms as suggested in Section~\ref{subsec:case2-qaoa}. The next step is to validate the same workflow on real quantum hardware, where calibration drift and nonstationary behaviors may further affect stopping rules and BHT's reliability.

\section{Threats to Validity}
\label{sec:threats}

\textbf{Internal validity} is influenced by implementation choices such as batch size, minimum-shot gate, pass-streak requirement, prior strength, and shot cap. Although these settings are fixed and reported explicitly, different parameter choices may affect stopping behavior and PASS rates, especially near the decision boundary. In addition, the matched fixed-budget baseline is only one comparator and should not be interpreted as exhausting the space of possible sequential or frequentist alternatives.

\textbf{Construct validity} is limited by the way correctness is operationalized in the two case studies. Both workloads reduce verification to a Bernoulli success predicate: correlated outcomes in Bell-state verification and Top-$K$ success in QAOA-MaxCut. These predicates are meaningful for the targeted tasks, but they do not capture all possible notions of correctness or quality.

\textbf{External validity} is limited because our evaluation uses \texttt{AerSimulator} and IBM fake backends rather than real quantum hardware. This improves reproducibility and allows controlled study of noise-model effects, but real devices may introduce additional variability, including calibration drift, queue delays, and nonstationary noise. Due to the limited access and the high cost of quantum hardware, further verification on real backends will be addressed in future research with more external support.

\textbf{Conclusion validity} is limited by the scope of the empirical study. The paper provides a proof-of-concept design study rather than a comprehensive benchmark over all verification strategies. Accordingly, our conclusions about the benefits of BHT are strongest for the tested settings and workloads, and the observed savings should be interpreted relative to the reported shot budgets and matched baseline configuration. Further validation on real hardware and broader benchmark suites is needed before drawing stronger claims.

\section{Conclusion and Future Work}
\label{sec:conclusion}

This paper presents Bayesian sequential verification as a practical workflow for quantum programs. Rather than relying on fixed-shot testing, the proposed approach uses reference-based priors, sequential posterior updates, and conservative one-sided LCB to make verification decisions under finite shot budgets. Across two complementary Qiskit-based case studies---Bell-state verification and QAOA Top-$K$ verification---this paper shows that the same verification logic transfers across qualitatively different success predicates and remains competitive against matched fixed-budget baselines.

Several directions remain for future work. First, prior selection and refresh should be systematized through lightweight prior management across repeated runs. Second, a variety of fault models could be explored when tests fail. Third, the same workflow should be validated on real quantum hardware. Together, these steps would move Bayesian sequential verification closer to reliable quantum software testing under tight measurement budgets.

\section*{Data Availability}

The Python scripts and experimental results are available at \href{https://doi.org/10.5281/zenodo.18341209}{\nolinkurl{doi:10.5281/zenodo.18341209}}.

\bibliographystyle{IEEEtran}
\bibliography{refs}

\end{document}